# Zero and negative energy dissipation at information-theoretic erasure

**Laszlo Bela Kish, Claes-Göran Granqvist, Sunil P. Khatri, Ferdinand Peper**

**Abstract** We introduce information-theoretic erasure based on Shannon's binary channel formula. It is pointed out that this type of erasure is a natural energy-dissipation-free way in which information is lost in double-potential-well memories, and it may be the reason why the brain can forget things effortlessly. We also demonstrate a new non-volatile, charge-based memory scheme wherein the erasure can be associated with even negative energy dissipation; this implies that the memory's environment is cooled during information erasure and contradicts Landauer's principle of erasure dissipation. On the other hand, writing new information into the memory always requires positive energy dissipation in our schemes. Finally, we show a simple system where even a classical erasure process yields negative energy dissipation of arbitrarily large energy.

_______________________________________________

L.B. Kish (✉) and S.P. Khatri
Department of Electrical and Computer Engineering, Texas A&M University, College Station, TX 77843-3128, USA
e-mail: laszlo.kish@ece.tamu.edu, sunilkhatri@tamu.edu

C.G. Granqvist
Department of Engineering Sciences, The Ångström Laboratory, Uppsala University, P.O. Box 534, SE-75121 Uppsala, Sweden
e-mail: claes-goran.granqvist@angstrom.uu.se

F. Peper
CiNet, NICT, and Osaka University, 1-4 Yamadaoka, Suita, Osaka, 565-0871, Japan
e-mail: peper@nict.go.jp

## 1 Introduction: Classical information erasure

In computer memories, the erasure of a bit means resetting its value to zero. This type of erasure, which we call classical erasure, implies a bit-value change if the bit value before the erasure was 1. In accordance with Brillouin's negentropy equation [1–3], any bit-value change gives a minimum dissipation of energy $E_d$ by

$$E_d \geq kT \ln\left(\frac{1}{p_e}\right) , \qquad (1)$$

where $p_e$ is the error probability of the operation $(p_e < 0.5)$, $k$ is Boltzmann's constant and $T$ is absolute temperature. In the case of $p_e = 0.5$, which is the limit for a completely inefficient operation, the relevant $kT \ln(2)$ dissipation is the famous Szilard–Brillouin–Landauer limit [1–3].

In this paper, we introduce "information-theoretic erasure", ITE, for which the elimination of information is guaranteed by information theory. We show that ITE does not cause energy dissipation, and it can even produce negative energy dissipation by cooling the environment. However, the writing of new information into the memory always requires positive energy dissipation.

## 2 Information-theoretic erasure

In accordance with Shannon's channel capacity formula for binary channels, the maximum information content (maximum mutual information between the input and output) $I_1$ of a single bit with error probability $p_e$ (error during transfer) is given by

$$I_1 = 1 + p_e \log_2 p_e + (1 - p_e)\log_2(1 - p_e) \qquad (2)$$

as illustrated in Fig. 1. The case of $p_e = 0$ yields an $I_1 = 1$ bit, while $p_e = 0.5$ corresponds to a random coin with an $I_1 = 0$ bit .

Motivated by these facts, we consider the memory an information channel between the Writer and Reader of information and introduce ITE as follows: Suppose that the bit-operations are error-free and that the bit-value before the erasure is 1. Then the probability $p(1)$ that the bit has the value 1 is

$$p(1) = 1 . \qquad (3)$$

Similarly, if the bit value before the erasure is 0, then the probability is

$$p(0) = 1 . \qquad (4)$$

We define *information-theoretic erasure* so that, after the erasure, these probabilities become

$$p(1) = 0.5 \quad \text{and} \quad p(0) = 0.5 , \qquad (5)$$

which guarantees total elimination of information from the memory. Alternative names such as "randomization", "thermalization", etc., could have been used, but we choose "information-theoretic erasure" which is inspired by the term "information-theoretic security" in communication [4]. In the latter case, the meaning is that—according to information theory—the information does not exist in the channel for the eavesdropper. Similarly, after ITE, and according to information theory (and Eq. 2), the information does not exist in the memory.

Finally, we note that Eq. 2 (and Fig. 1) does not only illustrate the zero-information case at $p_e$= 0.5 but is practically useful for evaluating incomplete erasure and remaining information.

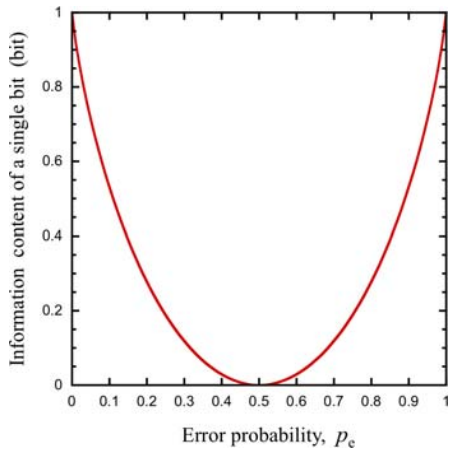

**Fig. 1** Information content of a bit versus error according to Eq. (2).

## 3 Physical realizations

In this section, we show two physical realizations: one is *passive* erasure (thermalization) in memories with double-potential wells and the other is *active* erasure in capacitor-based memories, where even negative energy dissipation is feasible.

*3.1 Passive erasure in symmetric potential wells*

The most natural process that leads to ITE is thermalization in a symmetric double-well potential system, such as in a magnetic memory; see Fig. 2. When such a system is kept untouched for a number of relaxation events, the exponential nature of relaxation will cause ITE so that

$$p(1) \to 0.5 \quad \text{and} \quad p(0) \to 0.5 \qquad (6)$$

occur without energy dissipation because equilibrium thermal fluctuations are utilized for erasure. Of course, such a process may take thousands of years, but the existence of this phenomenon proves that no energy dissipation is required for information erasure. Similar arguments may explain how the brain can easily forget neutral information, while the creation of new information requires effort.

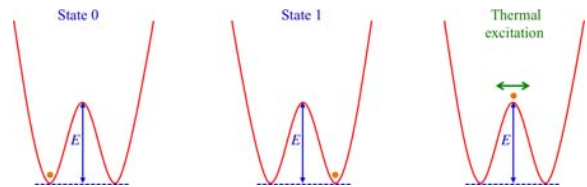

**Fig. 2** Passive information-theoretic erasure in a zero-energy-dissipation fashion by waiting for thermalization at ambient temperature, or, in a dissipative way, by heating the memory cell to rapidly thermalize the bit. Note that the information entropy in the memory is zero (see Sec.4) because the bit state is deterministic and exactly set/known by the operator.

Of course, it is possible to heat the memory cell so that $kT$ approaches the barrier height $E$ (see Fig. 2) sufficiently to cause rapid ITE, but this approach involves energy dissipation and is uninteresting from a fundamental scientific point of view.

*3.2 Charge-based bit with information-theoretic erasure*

We now consider a capacitor-type information cell. Figs. 3–6 show various aspects of its operation. Suppose that positive voltage is interpreted as bit 1 and negative voltage as bit 0.

Fig. 3 shows the writing process. An external resistor and voltmeter are connected to the cell, and thus a parallel *RC* circuit is present. As a consequence of the measurement and decision process described below, the writing process is strongly dissipative. The resistor will drive a Johnson noise current through the capacitor thereby yielding a noise voltage on the capacitor; see Fig. 4. The voltmeter monitors this voltage, and the resistor is disconnected when the required voltage is reached. The root-mean-square value of the Johnson noise voltage on a parallel *RC* circuit is [3]

$$\sigma = \sqrt{kT/C} \ , \tag{7}$$

and the corresponding mean energy in the capacitor is $kT/2$. Two cases should be considered:

(*i*) If, during the writing process, we choose $+\sigma$ for bit value 1 and $-\sigma$ for bit value 0 then the information-containing capacitor will possess thermal equilibrium energy in accordance with Boltzmann's equipartition theorem for a single thermal degree of freedom.

(*ii*) On the other hand, if we use the voltages $\pm u_0$ for the 1 and 0 bits, respectively, where $u_0 < \sigma$, then the energy in the capacitor will be *less* than the thermal equilibrium level $kT/2$.

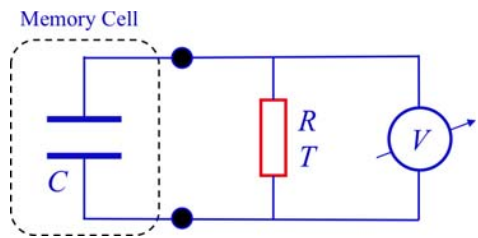

**Fig. 3** Writing of information into a capacitor by thermalization and measurement/control similar to that of electrical demons [3]. Johnson noise in the resistor drives the current, and the connection to the memory is terminated when the voltage level corresponding to the information to be stored is reached.

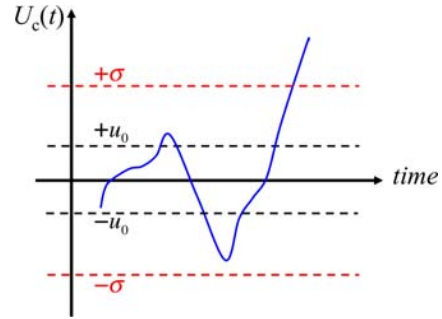

**Fig. 4** Johnson noise voltage in a capacitor. The voltage levels $\pm u_0$ pertain to written bit values; see the main text for details.

Fig. 5 shows the corresponding read-out process. It entails measuring the voltage and deciding if it is positive (bit value 1) or negative (bit value 0).

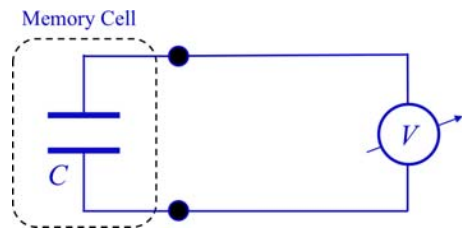

**Fig. 5** Reading out information from a capacitor. Here both the information entropy of the memory and the thermodynamic entropy related to the voltage are zero because the voltage is a deterministic quantity, which is exactly set/known by the operator (see Sec. 4). The system is isolated in a lower energy state than its environment by a deterministic, non-thermal, negative energy shift compared to the mean $kT/2$ thermal energy per thermal degree of freedom in the environment.

Fig. 6 illustrates the erasure process. The capacitor is reconnected to the resistor, but no voltage measurement or decision is necessary. The capacitor will be thermalized within a few events with relaxation time $\tau = RC$, and the conditions of relations (6) are reached. The energy dissipation during erasure is determined by our former choice:

(*iii*) Using writing condition (*i*), the mean energy of the capacitor will not change during erasure, and hence the energy dissipation is zero.

(*iv*) However using writing condition (*ii*), the mean energy of the capacitor is increased during information erasure so that energy dissipation is negative, and consequently the resistor and the environment of the memory cell are cooled in accordance with

$$E_{\text{diss}} = \frac{1}{2}\left(Cu_0^2 - kT\right) < 0 \quad . \qquad (8)$$

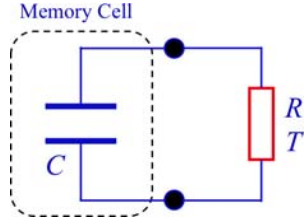

**Fig. 6** Information erasure by thermalization of a capacitor. For small values of $u_0$, energy is extracted from the resistor so that its environment is cooled, thus indicating negative energy dissipation. The voltage is a thermodynamic quantity representing thermal equilibrium. The information entropy of the memory cell is 1 bit (see Sec. 4).

**4 About some former theories**

There are some prior thermodynamic-theoretical investigations that mention both indeterministic operations and negative energy dissipation [5,6], where [6] is aiming to prove that logical reversibility allows physical reversibility, which we cannot accept. It is of interest to view the present work in context with these investigations. We believe that thermodynamic approaches to error-free memories are invalid, and our results and scheme are entirely independent from the considerations in earlier work [5,6].

We now give a brief summary of our opinion [7], which is not new in the literature as seen below. First we mention that Porod *et al*. [8,9], Porod [10] and Norton [11,12] have pointed out that the logic state in a computer is different from the system state in a thermodynamic ensemble, and its state-space is limited. It is possible to extend this discussion to include some further implications of determinism. The information entropy $S$ of a single-bit memory is given as

$$S = -k\left[p(0)\ln p(0) + p(1)\ln p(1)\right], \qquad (9)$$

where $p(0)$ and $p(1)$ are the probability of being in the state with bit values 0 and 1, respectively. Thermodynamic approaches assume that the information in the memory is statistical, which is incorrect and leads to a well-known flaw as discussed elsewhere [7].

However, for an error-free memory (a Turing machine), if the bit value is 0 then

$$p(0) = 1 \quad \text{and} \quad p(1) = 0 \qquad (10)$$

and, in accordance with Eq. 9 the corresponding entropy is

$$S_1 = -k\left[\ln(1) + 0\ln(0)\right] = 0 \qquad . \qquad (11)$$

Similarly, when the bit value is 1 one gets

$$S_1 = -k\left[0\ln(0) + \ln(1)\right] = 0 \quad . \qquad (12)$$

Therefore, in a deterministic computer (a Turing machine) with error-free memory, the information entropy of the memory is always zero. Furthermore, this fact is independent of the thermodynamic microstate of the memory, including the thermal motion of atoms, etc., in them. Thus thermodynamics could possibly be used only to interrelate information and entropy in memories with bit-errors caused by (thermal) fluctuations, which is not the topic of our paper or [5,6].

Such considerations are actually valid for the full deterministic Turing machine, not only for the memories. We reiterate that our observations are not new. For example, already Alfred Renyi [13] explained that the information in an article showing a new result of geometry is not more than the information represented by the axioms, the rules of calculations, and possible initial and boundary conditions. Beisbart and Norton [14] have similarly argued about Monte Carlo simulation results.

In conclusion, it is impossible to present our results in the context of earlier notions [5,6], which is a consequence of the lack of a common ground of understanding. However we may mention an earlier claim [5] that, in special cases, negative energy dissipation (cooling) is possible during classical erasure (when resetting bits to 0). This claim [5] is limited by

$$E_{\text{diss}} \geq -kT\ln(2)\Delta S^s , \qquad (13)$$

where $\Delta S^s$ is the change of the "information theoretic self-entropy of the erasable system" [5], i.e., in a single-bit memory, with 1 bit maximum entropy change. Under these conditions the cooling (negative) energy would be less than $kT\ln(2) \approx 0.6 kT$. However, as we have shown above, the information entropy (and its change) is zero during the whole operation of the error-free memory, and thus Eq. 13 gives zero energy (cooling or heating). More details on this matter are given below.

## 5 Conclusion and remarks

Our present study showed schemes and realizations for information erasure, for which energy dissipation can be zero or negative. We trust that these results lend further credence to objections [8–12,15,16] against the Landauer Theorem [17,18], which claims that information erasure is a dissipative process whereas information writing is not.

Following general practices [5,6,17,18] for analysis of information writing and erasure, we neglected the energy dissipation of the external control step for connecting the resistor to the capacitor in the case of erasure. Including the energy for this control [2,3] would imply positive net dissipation in the environment. However, the same happens also during information writing, and consequently information writing still comes out as much more dissipative than erasure.

Finally, we note that in a non-practical (non-electronic) fashion, it is also possible to introduce a memory with cooling during classical information erasure using the following scheme: From the above-described solutions we keep the principle of calling erased state (now the state 0) the state, which is in thermal equilibrium with the environment. Let us consider the specific case of an ice cube maker tray. The high bits are the cubes that are frozen, while the low bits are the water cubes thermalized at room (ambient) temperature. Erasure takes place when the ice cube is coupled to the ambient temperature. Obviously, the erasure will cool the environment. Due to the phase transition, this system is somewhat more in line with the thermodynamic approaches [5,6], but the information entropy of bits is still zero because of the deterministic values. Moreover, the cooling energy can be arbitrarily large depending on the volumes of ice cubes. This fact violates the limit given in the literature [5]; see Eq. 13 with the limit of $E_{\text{diss}} \approx 0.6\, kT$ obtained when we disregard the fact that the information entropy is always zero and assume the information entropy change of 1 bit in a single-bit memory. Instead, for a 10 cm$^3$ ice cube we have $E_{\text{diss}} \approx 10^{24}\, kT$, thus indicating the limitations of even the most advanced thermodynamic approach [5] to memories.

We observe that it is possible to swap the bit values of phases and assume that the ice phase is the bit value 0 and the water phase is 1. We can do this by assuming that the ambient temperature is below freezing. In this case, the erasure yields positive energy dissipation and the writing is cooling. However, such swapping of the bit value meanings is pointless in the information-theoretic erasure schemes above.